# Chaos in Liquid Crystal Directrons

Praveen Kumar Singh, Salman Ahmad Khan, **Soumik Das**\*

*Department of Chemical Engineering, IIT Kanpur, Kanpur, India - 208016.*

*\*Email:dsoumik@iitk.ac.in*

**Abstract:** Biological systems often operate at the boundary between order and chaos, transitioning from directed to irregular dynamics to achieve adaptability and robustness. Reproducing such transitions in artificial soft matter remains a central challenge. Here, we report a biomimetic regime of directron dynamics in achiral nematic liquid crystals, in which coherent, directed motion collectively evolves into chaos. Driven by multi-directron interactions, the system develops coexisting directron families with competing trajectories, displaying randomized motion, dynamic assembly formation and spontaneous fission of high energy to low energy daughter directrons - all of which mimics the phenotypic diversity observed in biological groups. Above a critical electric field, these interactions drive the system into a chaotic state that is distinct from the directed behaviours reported previously. We further introduce a minimal dipole-based model that qualitatively captures the underlying physics of this transition. Together, our results establish an artificial active system in which chaos emerges intrinsically from interactions, offering a versatile platform to study biological dynamics and opening new avenues for liquid-crystal-based soft-matter applications involving adaptive transport, cargo delivery, and energy transduction.

**Significance**: Living systems leverage chaos to efficiently explore energy landscapes, allowing them to be adaptable and resilient. However, reproducing chaos in artificial materials remains a challenge. This work demonstrates a minimal liquid crystal-based active system that can mimic biological chaos via propagating director deformations known as directrons. Through interaction-driven proliferation and spontaneous fission, directrons reproduce key chaotic signatures such as energy state exploration, collective disorder and growth-and-division. These findings establish LC-directrons as a versatile synthetic substitute for biological active matter and open pathways for designing new and efficient adaptive soft matter systems.

**Introduction:** A hallmark of biological complexity is that systems often transition from directed, predictable behaviour to chaotic, irregular dynamics as parameters such as feedback strength, coupling, or environmental interactions change[1–6]. The balance between order and chaos is essential in many biological contexts - order enables efficiency and reliability, while chaos provides adaptability, responsiveness and the ability to explore different energy states. Identifying new mechanisms by which such out-of-equilibrium systems can be artificially constructed is critical in the context of developing functional, intelligent, and adaptive soft matter.

Research carried out in the past five years have illustrated the formation and dynamic propagation of dissipative solitons within thin films of achiral nematic liquid crystals (LCs) when subjected to alternating current (AC) electric fields (EFs)[7,8,17,9–16]. Termed "director bullets" or directrons, these solitons represent spatially confined deformations of the LC director (manifested as bright localized domains) that can traverse substantial distances and retain their distinct identities after collision with another directron. Initial investigations revolved around confining a (-,-) LC between surfaces inducing uniform planar anchoring, effectively nullifying field-induced realignment and Carr-Helfrich instabilities within the LC film upon AC field activation[7,8]. While the applied EF stabilizes the LC's far-field orientation, the LC director within the directron undergoes oscillation mirroring the frequency of the electric field. This oscillation introduces out-of-plane LC distortions that break the fore-aft symmetry along the direction of propagation. In the absence of Carr-Helfrich instabilities, the formation of these dissipative directrons is attributed to flexoelectric polarization, arising from splay and bend LC distortions[18,19]. At lower frequencies (the conductive regime) and in LCs with positive dielectric anisotropy, free ions and electrohydrodynamic effects, including isotropic flows, are also thought to contribute to directron generation[10,11,15]. More recent studies have successfully demonstrated intriguing capabilities like localized energy transduction at deformable interfaces through phenomena like LC jetting and LC-directron-induced emulsification[20], directron-enabled accelerated colloidal transport in anisotropic media[21] as well as collective motion in directrons[14]. These works have highlighted the immense potential of directrons as a new class of active matter in liquid crystals, including mimicking several features of active biological systems.

Going beyond these past studies, herein, we reveal a biomimetic regime of directron dynamics in achiral nematic LCs, marked by a fascinating transition from coherent, directed motion to chaos. Like biological agents, this transition emerges from multi-directron interactions, leading to the coexistence of distinct directron families with competing trajectories. Additionally, our work demonstrates that high-energy directrons can spontaneously split into two low-energy daughter directrons with a splitting probability that increases with the applied EF. Beyond a threshold EF, the proliferation of these diverse trajectories drives the system to a chaotic state, which is fundamentally distinct from the directed behaviours reported previously.

**Results:** We performed experiments in which EF was applied to 5 μm thick films of achiral nematic LC 4'-butyl-4-heptyl-bicyclohexyl-4-carbonitrile (CCN-47) (Figure 1a). Beyond a certain applied field (e.g., 130 Hz, 33 V), we observe the formation of directron bullets that traverse in a direction parallel to the far-field LC anchoring at speeds of $100 - 200$ μm/s (Figure 1b). These directrons (referred as H-directrons here onwards) represent spatially confined deformations of the LC director that appear as bright localized domains within the LC medium and are similar in appearance to the high-frequency directrons reported previously [7,9].

However, our experiments reveal a different progression of LC deformations with EF than reported previously (Figure 1c). As the EF is increased, directrons are observed to swerve from their initial trajectories, parallel to the LC alignment, to angular paths. We refer to directrons that adopt angular paths as A-directrons. Fig 1d-e shows two such instances where an H-directron deviates by angles of 35° and 40°, respectively, from its initial trajectory when the applied voltage is 130 Hz, 35 V. These trajectory changes occur randomly with no spatial or temporal preferences, even in the absence of surface features such as dust particles or edges of electrodes, and multiple trajectories can originate from the same spatial location. The instantaneous propagation directions at 130 Hz, 35 V and their cumulative distributions over a period of 20 sec (Figures 1f-g) reveal a broad distribution of heading angles (defined as the magnitude of swerve from the initial horizontal trajectory). Figure 1g shows that while most directrons exhibit headings near 0° and $\pm 180°$ (corresponding to the H-state), a significant fraction displays intermediate angular headings.

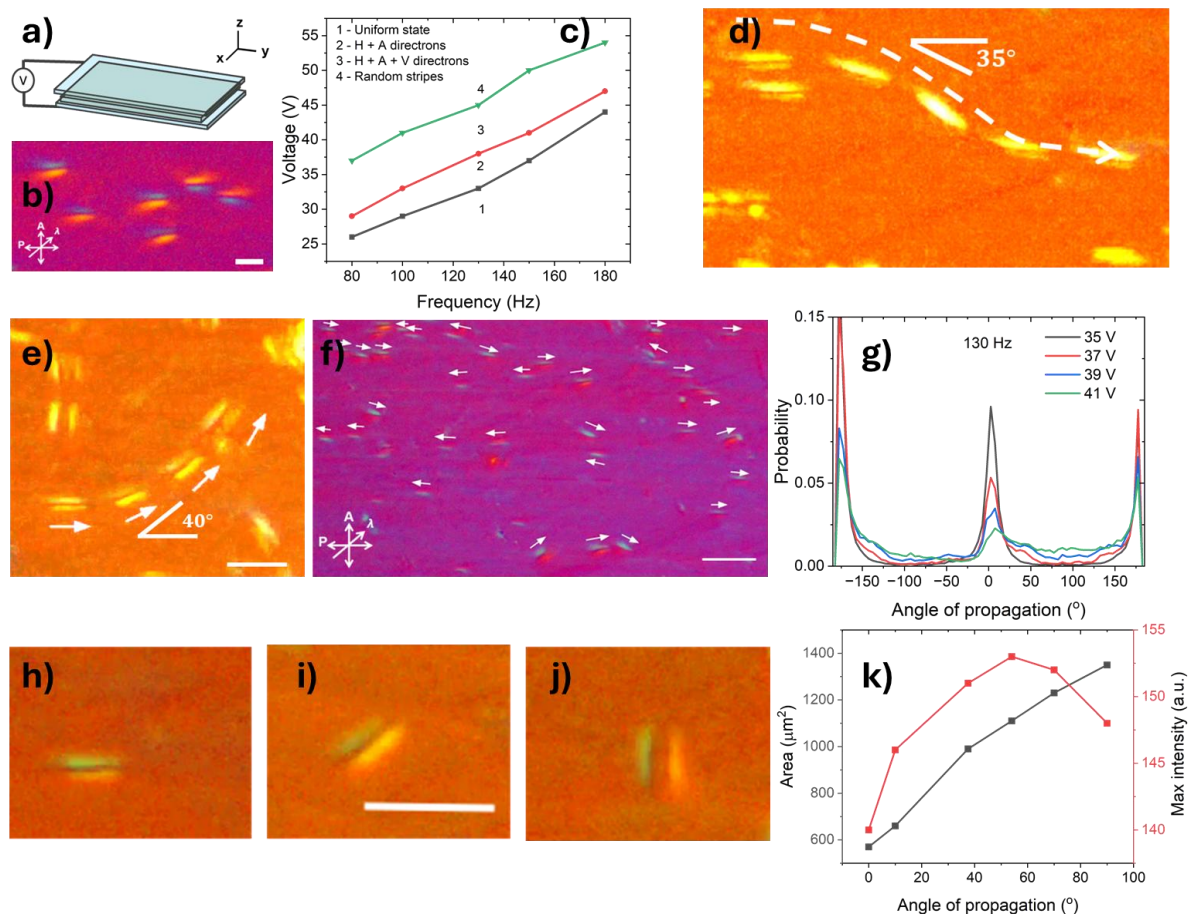

**Figure 1**. (a) Schematic illustration of the experimental setup used to study directrons. (b) Directrons formed at 130 Hz, 33 V. Images are taken under cross polars and a compensator. Scale bar 30 μm. (c) Voltage vs frequency phase diagram depicting the different regimes of directrons. (d-e) Time-stitched images of a propagating directron undergoing (d) 35° and (e) 40° swerves. Scale bar 50 μm. (f) Snapshots showing the instantaneous propagation direction of multiple directrons at 130 Hz, 35 V. Scale bar 100 μm. (g) Heading probability of directrons at different voltages. (h-j) Snapshots of (h) H, (i) A and (j) V-directrons observed under cross polars with a compensator. Scale bar 50 μm. (l) Variation of directron area and maximum intensity within a directron as a function of propagation direction.

The A-directrons possess internal structures similar to H-directrons (Figure 1h-j) and propagate at similar speeds. A closer examination, however, reveals that A-directrons are broader than the H-directrons (Figure 1k), indicating the presence of enhanced director distortions within – points we elaborate on later. These observations reveal, for the first time, the formation of high frequency directrons capable of propagating along multiple directions without time-varying EF or complex surface anchoring.

The presence of directrons travelling in multiple directions leads to new types of directron-directron interactions that can be head-on or angular, depending on the initial paths of the directrons. Figures 2a-d depict representative instances of anti-parallel and parallel interactions in our system. We observe that anti-parallel interactions generate either a deflection (Figure 2a) or scattering (Figure 2b) of the incoming directrons away from each other. The directrons deviate from their initial trajectories by approximately 20 − 40°, and then continue along the newly adopted heading. This behaviour is distinct from past reports, where the overall direction of propagation remains unchanged after collisions. Head-on encounters lead to directrons getting reflected in opposite directions. A single directron can undergo multiple such course changes, with each redirection determined by its encounters with other directrons.

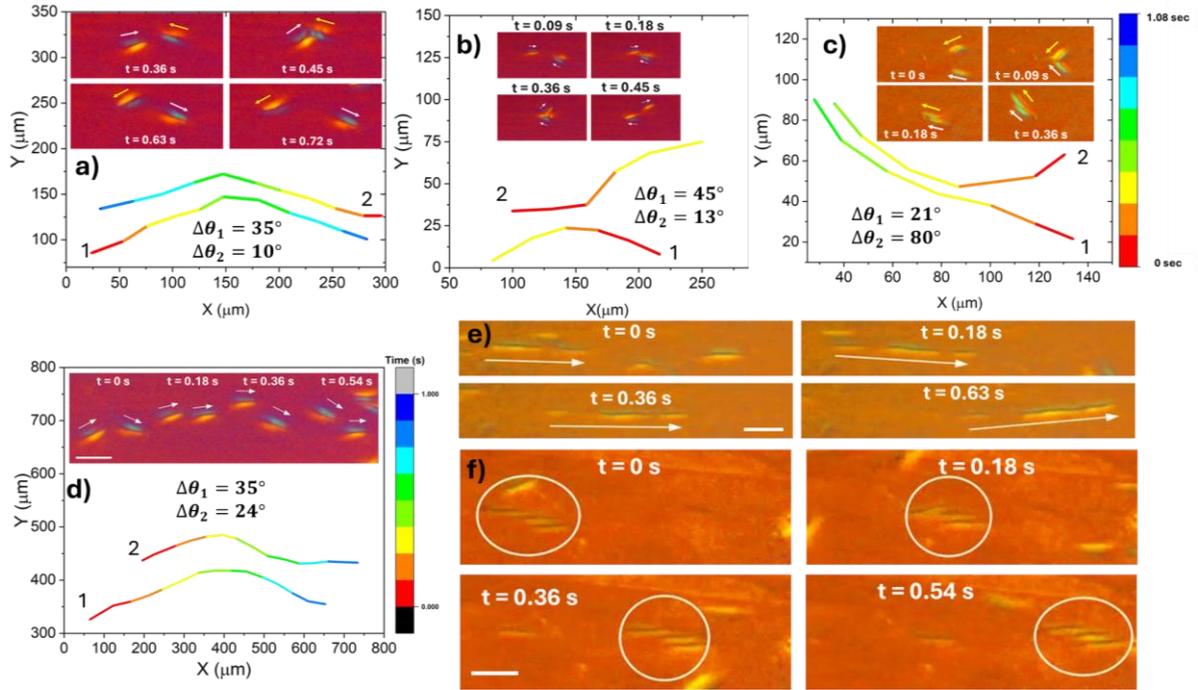

**Figure 2**. (a-d) Colour-coded cumulative trajectories of interacting directrons traveling in (a-b) similar and (c-d) opposite directions. Inset shows timed snapshots corresponding to the interactions. White arrows indicate the instantaneous direction of propagation. $\Delta\theta_1$ and $\Delta\theta_2$ represent the magnitude of swerves for directron 1 and 2 respectively. (e-f) Timed snapshots showing the formation of (d) directron chains and (e) directron stacks. Stacks are circled in white.

Along with head-on and angular interactions between opposing directrons, we also routinely observe a new class of interaction among directrons moving in the same overall direction (left-to-right or right-to-left) but with different headings. Parallel directrons travelling in the same direction, with identical headings, can assemble into end-to-end chains that travel together (Figure 2d). These directron chains can subsequently undergo the similar swerves observed for individual directrons (Figure 2d-e) Directrons with distinct headings exhibit oblique interactions leading to stacked configurations in which multiple directrons align and travel together (Figure 2c, f). These directron assemblies traverse significant distances, with varying headings, but may dissociate and adopt distinct trajectories in presence of other directrons. These observations demonstrate, for the first time, the formation and interactions involving such higher-order directron assemblies in achiral nematic LCs.

Interestingly, the A-directrons do not travel continuously along steady angular paths but instead sample a range of headings over their lifetimes (Figure 3a). These intermediate headings vary from one directron to another (Figure 3a). Moreover, A-directrons can also transition back and forth between H and A states after an interval of time that is determined by the applied EF (Figure 3a-b). Specifically, the average residence time in the A-state increases as the applied EF is increased (Figure 3b). At the same time, the number density of directrons increases significantly at higher EF, leading to more frequent directron-directron interactions and, consequently, additional trajectory changes, as discussed above. The combined influence of randomized angular headings, prolonged A-state lifetimes and enhanced directron interaction rates ultimately lead to the onset of chaotic directron motions at higher voltages.

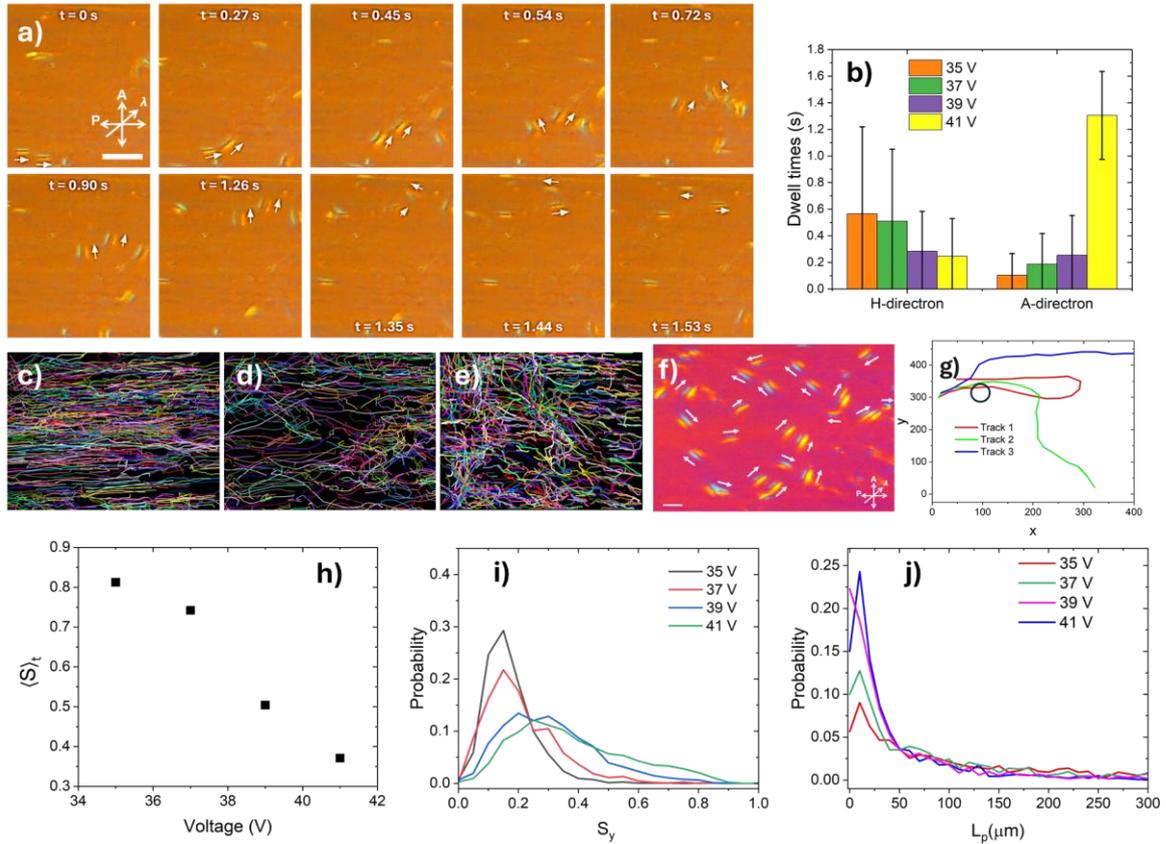

**Figure 3**. (a) Timed snapshots showing the spontaneous transition between H and A states in two directrons. White arrows indicate instantaneous propagation direction. Scale bar 100 $\mu m$. (b) Dwell time distribution for directrons at different voltages. (c-e) Cumulative tracks of directrons collected over an interval of 20s at (c) 130 Hz, 35 V, (d) 130 Hz, 37 V and (e) 130 Hz, 41 V. (f) Snapshots showing the instantaneous propagation direction of multiple directrons at 130 Hz, 41 V. (g) Trajectories of three distinct directrons which originate at the same spatial location (indicated by the black circle) but at different times. (h-j) Indicators of chaotic behaviour in directrons (h) nematic order, (i) Fraction of distance covered in y-direction and (j) persistence length. Error bars in (h) are smaller than the symbol sizes.

Figures 3c-f show the instantaneous and the combined trajectories of directrons captured over a 5s interval at different EFs. At lower EF (130 Hz, 35 V), most directrons exist in the H-state, and the swerve angles show a narrow spread around the horizontal direction (Figure 1g). In contrast, at higher EF (130 Hz, 41 V), the directrons show a greater tendency to exist in A-states, as observed from their instantaneous trajectories (Figure 3f). The distribution of swerve angles correspondingly broadens, reflecting a wide range of A-states (Figure 1g) - a high fraction of directrons now exhibit intermediate headings other than 0° and $\pm 180°$. The combined trajectories, collected over 5s (Figure 3e), further confirm the presence of multiple directron states with no preference in the overall direction of propagation. Figure 3g shows that three directrons originating from the same spatial location but observed at different times can adopt very distinct trajectories, highlighting the strong sensitivity of directron motion to initial conditions and subsequent interactions. Taken together, these observations confirm the transition from a directed motion of LC directrons, governed by the surface anchoring, to a chaotic behaviour that is determined by the applied voltage.

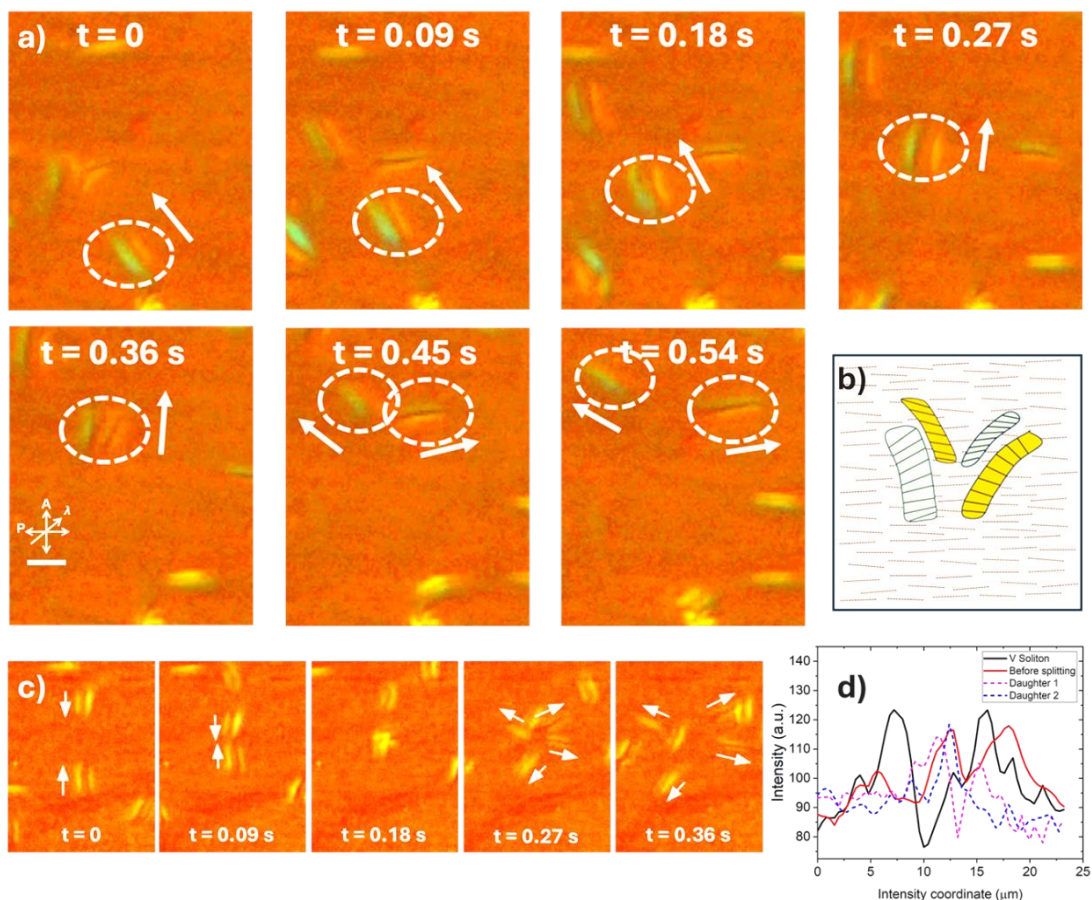

**Figure 4.** (a) Timed snapshots showing the splitting of a V-directron to two daughter directrons. (b) Schematic illustration of the state of the directron immediately after splitting. The dashed lines represent the LC orientation. (c) Timed snapshots showing the splitting of two colliding V-directrons into four daughter directrons. Scale bar 30 μm. (d) Intensity profiles of a V-soliton before and after splitting.

We confirm the transition to chaotic behaviour by additional trajectory analyses. As the applied voltage is increased, directron trajectories become increasingly irregular, exhibiting lower average nematic ordering, enhanced swerving, larger vertical displacements, and reduced trajectory coherence. Probability distributions of trajectory metrics show a clear voltage-dependent shift from predominantly directed motion at low voltages to strongly fluctuating dynamics at higher voltages (Fig. 3h–j). Collectively, these observations confirm a voltage-driven transition from directed to chaotic directron dynamics.

Among the A-directrons, the directrons that travel vertically at an angle of ±90° are designated as V-directrons. The intensity profile and the area of a V-directron is shown in Figure 1k-l, revealing that these directrons are wider than other directrons. In addition, their internal structure shows the presence of a thick deformation between the two directron arms (Figures 1j-k, 4a; $t = 0.27s$) that is not seen in other A or H-directrons. Remarkably, V-directrons are observed to spontaneously split into two daughter directrons, which adopt either A or H-states (Figure 4a). Time-resolved snapshots confirm that, immediately before splitting, the V-directrons further increase in size and the central deformation becomes more prominent (Figure 4a; $t = 0.36s$, Figure 4d). The split happens across this deformation (Figure 4b) and the resulting daughter directrons, having structures resembling A or H-directrons (Figure 4d), propagate in mutually opposite directions. The probability of splitting increases with applied

EF as V-state directrons become more frequent. Splitting events are often triggered by interactions with neighbouring directrons. Figure 4c illustrates such a scenario, where two V-directrons collide and then split into four A-state directrons that subsequently propagate in mutually opposite directions. The daughter directrons behave similarly to the other H and A-directrons and can themselves undergo further splitting later in their trajectories. The continuous splitting of directrons to daughter directrons with randomly distributed headings contributes to driving the chaotic dynamics observed in our system.

**Discussion:** The swerves in directron trajectories were observed over a substantially wide range of applied voltages and at high frequencies (100 – 200 Hz, Figure 1c). Very rarely, we could observe trajectory changes triggered by the presence of surface inhomogeneities (e.g., a dust particle). Moreover, we frequently observe instances where at the same location, a directron maintained its original trajectory while another directron swerved to an A-state. This eliminates the presence of surface anomalies as the origin of the observed chaotic directron behaviour. This intrinsic randomness of these trajectory changes has not been reported previously and is one of the central features underlying the onset of chaotic directrons in our system.

As the applied voltage is increased to 39 V and then 41 V, we see a significant increase in the number density of directrons. Unlike lower voltages, each directron now encounters multiple neighbouring directrons along its trajectory. We hypothesize that the elastic repulsion arising from proximity to other directrons induces frequent and random trajectory swerves throughout the system. Consistent to this, we observed that isolated directrons, without any neighbours within a radius of ~ 100 μm, can propagate in horizontal direction without any swerves even at high EFs. Similarly, A-directrons can also travel in a consistent angular trajectory in the absence of nearby directrons. A direct correlation between dwell times of A-directrons and directron densities, further supports this interaction-driven mechanism of trajectory changes. The splitting of V-directrons into two A/H-directrons also amplifies the directron density and the directron-directron interactions at higher EFs. These factors eventually trigger the transition from a primarily directed motion to a regime of chaotic directron dynamics. Under these conditions, the influence of the far-field LC anchoring becomes largely immaterial, and the directrons are free to adopt a range of A-states.

Our system uses the achiral nematic LC CCN-47 without any added ions or dopants. In the absence of such dopants, we attribute the onset of chaotic behaviour to a combination of directron-directron interactions and a weak surface anchoring. The surfaces were spin-coated with a 1 wt% PVA solution followed by gentle unidirectional rubbing. We estimate the anchoring energy to $\sim 10^{-6} - 10^{-5}$ J/m$^2$, and we hypothesize that the weak surface anchoring allows the directrons to sample multiple energy states associated with distinct propagation directions, rather than being constrained to a single direction dictated by the far-field LC anchoring. Indeed, as we increased the anchoring strength by increasing the PVA concentration to 3 wt% and above, we recovered the directed behaviour of directrons. As discussed below, a non-equilibrium steady-state analysis and a multi-particle simulation model also support this hypothesis.

The LC deformations are more pronounced in A-state directrons than in H-state directrons, as evidenced by their larger widths (Figure 1k-l). These deformations are further amplified as the direction of propagation approaches the vertical direction (Figure 1l), confirmed by the presence of an additional central band, of width ~5μm, between the two arms of the directron. The analysis above suggests that the different classes of directrons can be viewed as distinct energy states within an overall energy landscape $\Lambda(d; U) = \Lambda_0(d) - eUd$ (Figure 5a), where $d$ is the directron thickness, $\Lambda_0(d)$ represents the elastic and anchoring penalties and $-eUd$ represents the flexoelectric stabilization. At low EFs, just above the threshold $U_0$, H-state is the lowest energy configuration. The thicker angular directrons incur greater anchoring and elastic penalties and are energetically prohibited. Higher EFs enhance flexoelectric stabilization, progressively lowering the energetic cost of A-states and biasing the landscape in their favour (Figure 5a).

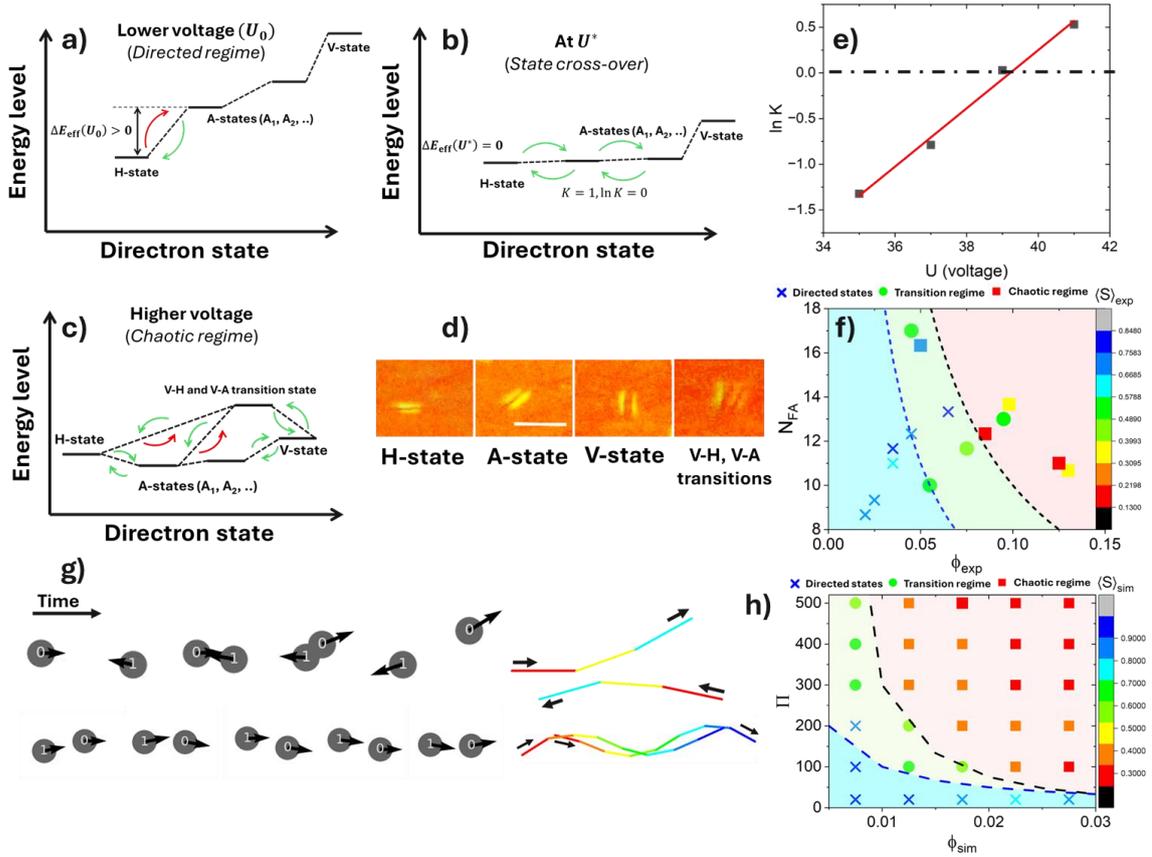

**Figure 5.** (a-c) Proposed energy landscape to explain the directron transitions and onset of chaotic directron dynamics as the applied voltages are increased. Green arrows denote allowed transitions. Red arrows indicate inaccessible moves. $A_1, A_2$ represent different angular states. (d) Optical micrographs of different energy states. (e) Plot of $\ln K$ vs $U$. Red line shows the data fitted to a straight line (f) Phase diagram, showing the dependence of $\langle S \rangle_t$, obtained from experiments. Black dashed curve in (f) represents $N_{FA}\phi_{exp} = 1$. (g) Simulated pair-wise interactions between anti-parallel and parallel directrons. (h) Phase diagram, showing the dependence of $\langle S \rangle_t$, obtained from simulations. The blue dashed curve in (h) represents $\Pi\phi_{sim} = 1$.

Importantly, H-A transitions are not accessed spontaneously but by overlap of LC-deformation fields during directron-directron interactions. Weak far-field anchoring permits these A-directrons to persist for a finite time during which they can explore multiple A-states through repeated interactions (Figure 3a). Increasing the applied voltage simultaneously increases directron speed and number density, thereby enhancing the interaction frequency. The steady-state population of angular directrons is thus controlled by a competition between interaction-driven excitation and anchoring-controlled relaxation. At lower voltages, where interactions are minimum, directron behaviour is directed and dictated by the far-field LC anchoring. At higher voltages, directron density, directron-directron interactions and flexoelectric stabilization are enhanced, which collectively promotes angular trajectories and an eventual transition to chaotic dynamics.

We can describe these events by a kinetic balance between the rate of interaction-induced creation of A-directrons and their relaxation back to H-states. The resulting population ratio can be written in an effective Arrhenius-like form $\ln K = -\Delta E_{\text{eff}}/E_{\text{int}}$, where

$$\Delta E_{\text{eff}}(U) = \Lambda(d_A; U) - \Lambda(d_H; U) = \Delta E_{\text{eff}}(U_0) - e\Delta d(U - U_0), \qquad U \geq U_0$$

Here, $d_A$ and $d_H$ are the thicknesses of A and H-directrons respectively, and $\Delta E_{\text{eff}}(U_0) = \Delta \Lambda_0 - eU_0\Delta d$. At the directron nucleation threshold $U_0$, flexoelectric stabilization is sufficient to stabilize H-directrons but not A-directrons, such that $\Delta E_{\text{eff}}(U_0) > 0$. The effective fluctuation scale $E_{\text{int}}$, arising from directron-directron interactions, is expected to vary weakly with EF, leading to an approximately linear dependence of $\ln K$ on $U$ along with the presence of a crossover voltage $U^* = U_0 + \Delta E_{\text{eff}}(U_0)/e\Delta d$, above which A-state directrons are increasingly favoured. We analyzed more than 4000 trajectories over a 20s interval and the estimated values of $K$ as a function of $U$ are shown in Figure 5b. The steady-state population ratio $N_A/N_H$ increases with applied voltage, indicating growing stability and population of A-directrons at higher EFs. We further confirm that $\ln K$ depends linearly on $U$, and changes sign at a finite voltage $U^* \approx 39$ V, indicating a crossover from H-state dominated to A-state dominated directron populations. Our estimates yield $U^* - U_0 \approx O(1 \text{ V})$, in good agreement with observations.

Physically, the crossover at $U^*$ marks the voltage at which EF-induced stabilization of A-directrons overcomes the combined elastic and anchoring penalties. This crossover ($U^* = 39$ V) also coincides with the onset of enhanced orientational switching and complex spatiotemporal dynamics ($\langle S \rangle_t < 0.5$, Figure 3h), indicating that the emergence of chaotic behaviour is directly linked to the change in effective energy landscape of the directron states – the energy minima associated with H and A-states become comparable. The LC anchoring energy determines the relative ease with which distinct directron states can be sampled – a high surface anchoring energy can inhibit the appearance of these distinct directron states, and the system can directly transition to space-filling stripes, as was reported previously, and also observed in our systems when experiments were performed with confining surfaces that have been spin-coated with higher concentrations of PVA. We combined results, obtained at different EFs and surface anchoring, and prepared a phase diagram governing the directron dynamics (chaos metric $\langle S \rangle_t$) in terms of two control parameters: the ratio of flexoelectric to surface anchoring energy $N_{FA}$, which sets the lifetime of A-directrons, and the area fraction associated with directron deformations $\phi_{\text{exp}}$, which controls the interaction frequency (Figure 5d). The transition from directed to chaotic dynamics is predicted to occur when the product of these parameters exceeds unity. A reasonable agreement can be seen from Figure 5d.

To further validate the proposed mechanism of directed to chaotic transition in directrons, we developed a multi-particle interaction model that adopts a minimal set of physics observed in our experiments. Each directron is represented as a self-propelled point dipole moving in a two-dimensional plane with periodic boundary conditions. Particles propagate at a constant speed and carry a dipole moment that varies with the direction of propagation, lowest for H and highest for A-states. Dipole–dipole interactions are computed using a long-range force law, reproducing attraction and repulsion between parallel and anti-parallel directrons respectively. Short-range steric encounters are handled through a head-on collision rule that reverses particle velocities upon approach within some characteristic distance. To capture surface-anchoring effects observed experimentally, vertical motion is damped by a linear restoring force proportional to the vertical velocity component. The equations of motion are integrated explicitly in time under a constant-speed constraint, characteristic of active particle systems[22]. The resulting dynamics are governed by a single dimensionless control parameter $\Pi = \mu_V^2/(\gamma a_0^3 v_0)$, which captures the competing effects of voltage-induced stabilization of angularly moving particles and the surface anchoring effect that dampens oblique trajectories. The chaotic behaviour of directrons is finally quantified using a time-averaged nematic order parameter.

Our simple model can successfully replicate the experimentally observed pair-wise behaviours during parallel and anti-parallel directron interactions (Figure 5c). Subsequently, we performed the simulations for different values of $\Pi$ and particle densities $\phi_{sim}$ and obtained a phase diagram that is qualitatively similar to experimental observations, and clearly indicates a transition to chaotic behaviour as $\Pi$ and $\phi_{sim}$ are increased (Figure 5e). A more comprehensive model, that takes into account the complex LC interactions between finite-sized directrons is necessary for a deeper understanding of this phenomenon. Even then, our minimal model confirms our hypothesis that the underlying competing effects of flexoelectric stabilization and surface anchoring governs the seemingly complex and chaotic directron behaviours.

While splitting events have been reported in the past[9], they were rare, confined to LC domain boundaries, and their underlying mechanism remained unclear. The soliton fractalization reported by Aya and Araoka was a synchronized process in which all resulting solitons exhibited identical, directed motion[13]. In contrast, our work reports, for the first time, consistent observations in which randomized directrons are spontaneously and continuously generated from existing directrons. At high voltages, V-directrons, which exhibit the highest LC deformations, grow in size and develop complex changes in its internal structure (Figure 4 a, d, inset of Figure 5a). Splitting, therefore, provides a relaxation pathway from these high-energy transient intermediates to low energy A and H-states (Figure 5a). This process is favoured at higher voltages where the frequency of formation of V-directrons and the probability of transitions are higher. The observation that many of these splitting events happen because of proximity to other directrons further confirms the role of directron-induced perturbances as the fluctuation source in our system. The overall picture is thus that of a complex energy landscape explored by the directrons either by directly transitioning from one energy state to another or via splitting into two lower energy states (Figure 5a). The coexistence of multiple relaxation pathways underlies the emergence of chaotic dynamics in this system.

Several observations distinguish these results from past studies on LC directrons. First, in contrast to the complex strategies such as spatially patterned anchoring[9,23] or time-dependent EF[8], used previously to generate spatiotemporal responses in directron behaviour, our system employs only a uniformly weak surface anchoring. This minimal condition is sufficient to generate electrically tuneable distribution of horizontal, vertical and obliquely traveling directrons coexisting at the same time and at the same location (at a different time). Second, the weak surface anchoring enables new classes of oblique and parallel directron encounters, that generates randomized directron trajectories and the formation of higher-order assemblies such as directron chains and stacks. Such interactions were inherently absent in earlier systems, where directrons propagated along a single direction with uniform velocities. These non-standard directron interactions are central to the eventual transition to chaotic directron states, not reported previously. Importantly, chaos in the system emerges from interaction-stabilized exploration of different energy states, rather than externally applied noise or feedback, and this is qualitatively captured by our non-equilibrium steady state analysis and a minimal dipole-based directron model. Finally, our experiments help unmask the conditions underlying directron splitting. While previous studies have emphasized the need for nucleation sites to generate directrons[19], the splitting mechanism reported here ensures a continuous production of directrons, from existing directrons, with distinct energies and trajectories. Overall, our work reveals a previously unexplored regime of interacting directrons that collectively transition from directed motion to chaos.

Our work also significantly expands the current capabilities in mimicking the complex chaotic behaviour of living systems. The spontaneous swerves in directron trajectories closely parallel the bacterial run-and-tumble motion. More broadly, the chaotic behaviour of directrons resemble microorganisms continuously sampling distinct energy states, a capability that is crucial for their survival under hostile conditions. We also report the formation of higher order directron structures that resemble chaotic behaviour in groups of microorganisms. The unique observation of directron splitting mimics biological growth-and-division processes wherein microorganisms first increase in size (via energy/nutrient uptake) before undergoing fission. Together, these artificial yet self-organized chaotic behaviours establish directrons as a versatile proxy to study dynamics in living systems, with implications for LC-based soft matter technologies such as accelerated transport, colloidal cargo delivery and focussed energy transduction.

**Acknowledgements**

This work was supported by the Prime Minister's Early Career Research Grant (PM-ECRG) of Anusandhan National Research Foundation (ANRF), Government of India (Grant No. ANRF/ECRG/2024/004324/ENS-G), and Indian Institute of Technology Kanpur under the Initiation Grant scheme (Grant No. IITK/CHE/2023109). P.K.S. gratefully acknowledges support from the Institute Postdoctoral Fellowship of Indian Institute of Technology Kanpur.